%% file: wegg-arxiv-cleaned.tex
\documentclass[10pt]{article}

\usepackage[dvips,letterpaper,top=0.5in, bottom=0.5in, left=0.75in, right=0.5in,includefoot,heightrounded]{geometry}

\usepackage{srcltx}

\usepackage{amsmath}
\usepackage{amssymb,amsfonts}

\usepackage{abstract}

\usepackage{epsfig}
\usepackage[usenames,dvipsnames]{color}
\usepackage[usenames,dvipsnames]{xcolor}
\usepackage{subfigure}

\usepackage{xspace}
\usepackage{booktabs}

\usepackage{flushend}

\usepackage{cite}
\usepackage{url}

\usepackage{multirow}

\usepackage[short,12hr]{datetime}
\usepackage[amssymb]{SIunits}

\usepackage{hyperref}

\usepackage[sc,tiny]{titlesec}
       \usepackage{mathptmx}



\title{%
Optimal Wavelets for Electrogastrography}

\author{%
R.~J.~Cintra${}^{\text{1,2}}$,\thanks{This work was
supported by the Natural Sciences and Engineering Research Council
of Canada (NSERC) and the National Council for Scientific and
Technological Development (CNPq, Brazil).}
I.~V.~Tchervensky${}^{\text{2}}$,
V.~S.~Dimitrov${}^{\text{2}}$,
M.~P.~Mintchev${}^{\text{2,3}}$\\
{\normalsize ${}^{\text{1}}$Department of Electronics \& Systems, Federal University of Pernambuco, Recife, Pernambuco, Brazil}\\
{\normalsize ${}^{\text{2}}$Department of Electrical and Computer Engineering, University of Calgary, Calgary, Alberta, Canada T2N~1N4}\\
{\normalsize ${}^{\text{3}}$Department of Surgery, University of Alberta, Edmonton, Alberta, Canada T6G~2B7}}

\author{%
R. J. Cintra%
\thanks{%
R. J. Cintra was
with
the
Low-Frequency Instrumentation Laboratory, University of Calgary,
Canada.
Currently he is with 
the Signal Processing Group,
Departamento de Estat\'istica, 
Universidade Federal de Pernambuco.
Email:~\url{rjdsc@de.ufpe.br}}
\quad
I.~V.~Tchervensky%
\thanks{
I.~V.~Tchervensky was
with
the
Low-Frequency Instrumentation Laboratory, University of Calgary,
Calgary, Alberta, Canada T2N~1N4.}
\quad
V.~S.~Dimitrov%
\thanks{%
V.~S.~Dimitrov
is
with
Department of Electrical and Computer Engineering, University of Calgary, Calgary, Alberta, Canada T2N~1N4.}
\quad
M.~P.~Mintchev%
\thanks{%
M.~P.~Mintchev
is
with
Department of Electrical and Computer Engineering, University of Calgary, Calgary, Alberta, Canada T2N~1N4
and
the Department of Surgery, University of Alberta, Edmonton, Alberta, Canada T6G~2B7.
Email:~\url{mintchev@ucalgary.ca}
}
}

\date{\today\ @ \currenttime}
\date{}

\newcommand{\myabstract}{%
Matching a wavelet to class of signals can be of interest in feature
detection and classification based on wavelet representation.
The aim of this work is to provide a
quantitative approach to the problem of matching
a wavelet to electrogastrographic~(EGG) signals.
Visually inspected EGG recordings from sixteen dogs and six volunteers
were submitted to wavelet analysis.
Approximated wavelet-based versions of EGG signals were calculated
using Pollen parameterization of 6-tap wavelet filters and
wavelet compression techniques.
Wavelet parameterization values that minimize the approximation error
of compressed EGG signals were sought and considered optimal.
The wavelets generated from the optimal parameterization values
were remarkably similar to the standard Daubechies-3 wavelet.
}

\newcommand{\mykeywords}{%
Electrogastrography, gastric electrical activity, matching wavelets, optimization techniques.
}

\newcommand{\egg}{EGG\xspace}
\newcommand{\prd}{\operatorname{PRD}}
\newcommand{\CR}{\operatorname{CR}}

\begin{document}

\twocolumn[%
  \maketitle
  \begin{onecolabstract}
    \myabstract
  \end{onecolabstract}
  \begin{center}
    \small
    \textbf{Keywords}
    \linebreak
    ~\vspace{1cm}
    \mykeywords
  \end{center}
  \bigskip
]
\saythanks

\section{Introduction}

Cutaneous recordings of gastric electrical activity (GEA)
known as electrogastrography (\egg),
can play a major role in the  diagnosis
of gastric motility disorders~\cite{Smout80}.
Because of its low-cost and
non-invasiveness, the \egg technique has great appeal as a
clinical tool
and has been related to various gastric motility abnormalities~\cite{Borto98}.
Multiple studies have been conducted in other to analyze \egg recordings.
Although signal processing of \egg signals has been considered essential for
extracting clinically relevant information,
various traditional methods have been utilized with limited success~\cite{Verhagen:1999}.

Recently, advanced signal processing techniques, such as
wavelets, have been employed to analyze
electrogastrograms~\cite{Xie98,Ryu:02,Liang96,Liang02,Qiao96}.
This approach has been used to
(i)~propose new wavelets that can offer a better time-frequency localization of
\egg recordings~\cite{Xie98,Ryu:02};
(ii)~perform noise detection in \egg signals~\cite{Liang96};
(iii)~cancel artifacts related to stimulation~\cite{Liang02};
and
(iv)~characterize global gastric electrical dysrhythmias~\cite{Qiao96}.
An important aspect of wavelet analysis
is related to designing a wavelet that matches a class of signals.
Although wavelet matching can be of great importance
for detection and classification~\cite{Chapa2000},
wavelets that match \egg signals have not been systematically sought.

The present study addresses the problem of finding a wavelet that best ``matches''
the waveshape of \egg signals in basal state.
Although there are numerous issues concerning
the choice of wavelet for signal analysis~\cite{Chapa2000},
generally,
a wavelet can be regarded as best suited to a class of signals
if the latter can be represented by
as few wavelet coefficients as possible~\cite{Mallat:03,Vetterli:2001}.
Thus, wavelets which resemble the waveshape of the signal under analysis are often selected.

In the framework of the proposed research methodology,
an optimal wavelet is sought that can adequately
represent a wavelet-compressed \egg signal at a given
compression ratio.
The optimality is detected by minimizing
an error measure between the original signal and its
compressed version, subject to the choice of wavelet.
If,
 for a given wavelet,
   the error associated with the compressed signal were minimal,
then
  its wavelet coefficients were considered to best represent the original signal.
Therefore,
the selected wavelet would more effectively ``match''
the signal under analysis when compared to other wavelets in consideration~\cite{Gopi1994}.

Consequently, the ultimate aim of this study is to quantitatively determine
a wavelet suitable for the analysis of basal \egg recordings in canine and human models.

\section{Methods}

\medskip

\subsection{Experimental Setup}

\subsubsection{Canine Experiments}

\begin{figure}
\centering
\subfigure[]{\includegraphics[scale=0.85]{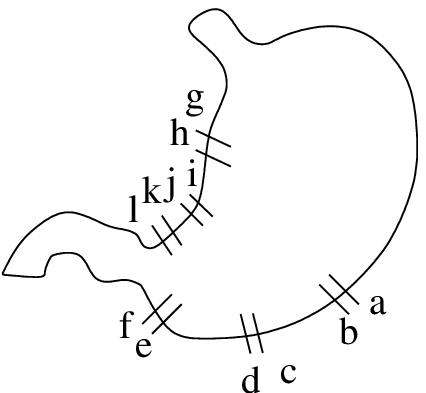}}
\subfigure[]{\includegraphics[scale=0.85]{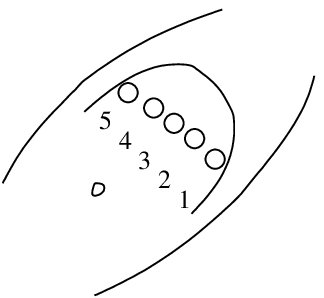}}
\\
\subfigure[]{\input{electrode_table_gea.tex}}
\quad
\subfigure[]{\input{electrode_table_egg.tex}}
\caption{Internal~(a) and cutaneous~(b) electrode positioning in canine experiments.
Various electrode combinations were used for the GEA~(c) and the EGG~(d) recordings.
The electrode combination for the EGG recordings in human experiments was similar.}
\label{egg:pos}
\end{figure}

After a laparotomy and the installment of six pairs of internal
subserosal stainless steel wire electrodes into the antral gastric
wall of sixteen acute dogs (seven female and nine male),
the abdominal wall was closed and five neonatal
electrodes (Conmed, Andover Medical, Haverhill, MA, USA)
were placed cutaneously along the projection of the gastric axis.
These five electrodes were grouped to form eight \egg channels.
In addition,
the setup provided six internal GEA channels.
However, in the present study only the \egg channels were processed,
while
the internal GEA recordings were used as a visual reference only
to verify that normal electrical activity was present.
The electrode combination set and
a diagram of the physical location of the electrodes are depicted
in Figure~\ref{egg:pos}.

Thirty-minute \egg recordings were performed in the basal state.
The captured \egg  signals were conditioned by a 0.02--0.2~\hertz\
low-pass
first order Butterworth active filter.
After amplification, 12-bit analog-to-digital conversion was performed
using a sampling frequency of 10~\hertz\
and
{\sc Labmaster 20009} 16-channel analog-to-digital converter
(Scientific Solutions, Vancouver, BC, Canada).

\subsubsection{Human Experiments}
Using a similar 8-channel \egg configuration,
one-hour recordings from six normal volunteers (two female, four male)
in postprandial state (500~Kcal, 52\% carbohydrates, 19\% proteins, and 29\% fat) were obtained.
The average body mass index for the volunteers was~22.2~\kilogram\!\cdot\!\rpsquare\metre\
(SD 3.0~\kilogram\!\cdot\!\rpsquare\metre).
Signal conditioning,
amplification and digitization process similar to
the ones utilized in the canine experiments were
implemented.

All experiments were approved by the Animal Welfare Committee and
the Ethics Committee at the Faculty of Medicine, University of Alberta.

\subsubsection{Signal Preprocessing}

Since both canine and human recordings were of significant
duration,
the raw \egg data
were intermittently contaminated with a multitude of artifacts,
including:
(i) motion artifacts;
(ii) spontaneous variations in electrode potentials;
(iii) respiration;
(iv) signal saturation  during recording;
(v) electrocardiac activity;
and
(vi) loss of signal during recording.
Usually these artifacts appeared simultaneously in all recording channels.
Some of these noisy patterns were visually evident (e.g., iv and~vi)
and could be easily identified and
discarded~\cite{Verhagen:1999}.
This practice has been recommended before in order to obtain a more reliable signal for subsequent
analysis~\cite{Parkman:03}.

Therefore, for each subject, a 10-minute time interval %
of channel-synchronized data
was manually selected.
These data were considered to be free from identifiable noise patterns.

\subsection{Signal Analysis}

\subsubsection{Wavelet Compression}

In a discrete-time formalism
wavelet transforms are performed via the Fast Wavelet Transform using
Mallat's pyramid algorithm for decomposition (forward transform) and
reconstruction (inverse transform)~\cite{Mallat:03}.

\begin{figure}
\centering
\epsfig{file=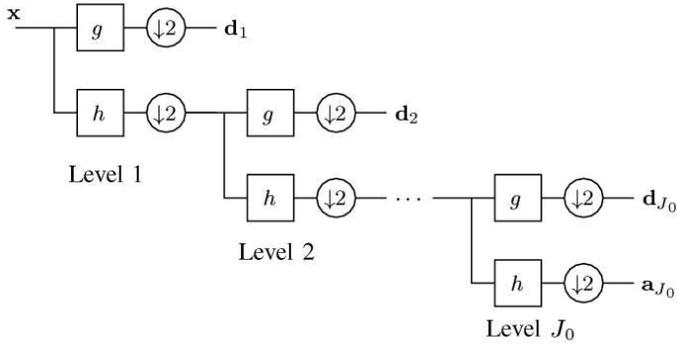,width=1\linewidth}
\caption{Wavelet analysis filter bank. The signal is iteratively decomposed through
a filter bank to obtain its discrete wavelet transform.}
\label{filter_bank}
\end{figure}

Let $\mathbf{x}$ be a discrete signal with $N=2^J$ points
(a sampled version of the analog signal $x(t)$).
The discrete wavelet transform (DWT) of $\mathbf{x}$ is computed in a recursive
cascade structure consisting of
decimators $\downarrow\!\!2$
and
complementing filters $h$ (low-pass) and $g$ (high-pass),
which are uniquely associated with a wavelet~\cite{MisiMisi00}.
Fig.~\ref{filter_bank} depicts a diagram of the filter bank structure.

At the end of the algorithm computation,
a set of vectors is obtained
$\{ \mathbf{d}_1, \mathbf{d}_2, \ldots, \mathbf{d}_j, \ldots, \mathbf{d}_{J_0}, \mathbf{a}_{J_0} \}$,
where~$J_0$ is
the number of decomposition scales of the DWT.
This set of approximation and detail vectors represents the DWT of the original signal.
Vectors~$\mathbf{d}_j$ contain the DWT detail coefficients of
the signal
in each scale~$j$.
As $j$ varies from~1 to $J_0$,
a finer or coarser
detail coefficient vector is obtained.
On the other hand, the vector~$\mathbf{a}_{J_0}$
contains the approximation coefficients of the signal at scale~$J_0$.
It should be noted that this recursive procedure can be iterated~$J$ times at most.
Usually, the procedure is iterated~$J_0< J$ times.
Depending on the choice of~$J_0$, a different set of
coefficients can be obtained.
Observe that the discrete signal~$\mathbf{x}$ and its DWT have the same length~$N$.
The inverse transform can be performed using a similar recursive approach~\cite{Mallat:03}.
Generally, a signal can be subject to various wavelet decompositions.
The analysis depends on
(i) the choice of wavelet (filters $h$ and $g$);
and
(ii) the number of decomposition levels (scales)~$J_0$.

A wavelet-based compression scheme aims to satisfactorily represent
an original discrete signal~$\mathbf{x}$ with as few DWT coefficients
as possible~\cite{Vetterli:2001,Chagas:00,Lu:99}.
One simple and effective way of doing that is to discard
the coefficients that, under certain criteria, are considered
insignificant.
Consequently, the signal reconstruction
is based on a reduced set of coefficients~\cite{Vetterli:2001,Unser1996}.

In the present work the classic scheme for non-linear compression
was used~\cite{Vetterli:2001}.
This procedure considers an \emph{a posteriori} adaptive set,
which keeps  $M$
wavelet transform coefficients
that have the largest absolute values.
A hard thresholding was used to set the remaining coefficients to zero.
The number of coefficients $M$ to be retained was determined according
to the desired compression ratio $\CR$, which was defined by
\begin{equation}
\CR = \frac{N}{M},
\end{equation}
where $N$ and $M$ are the number of wavelet transform coefficients of
the original and the compressed signals, respectively.

\subsubsection{Measurement of Distortion}

To further the analysis, it is necessary to introduce
an error measure %
to compare the original discrete signal~$\mathbf{x}$
with its reconstruction~$\tilde{\mathbf{x}}$.
Several measures that allow the
evaluation of the effect of compression schemes
have been suggested~\cite{Besar:00}.
However, one of the most commonly used is
the Percent Root-mean-square Difference
($\prd$)~\cite{Chagas:00,Besar:00,Lu:99},
which was utilized in the present study
as a measure of distortion
in the compression scheme.
The $\prd$ of two signals, $\mathbf{x}$ and $\tilde{\mathbf{x}}$,
both of length~$N$,
is defined by:
\begin{equation}
\prd(\mathbf{x},\tilde{\mathbf{x}})
=
\sqrt{\frac{\sum_{i=0}^{N-1}  (x_i - \tilde{x}_i)^2 }{\sum_{i=0}^{N-1} x_i^2}}
\times
100 \%.
\end{equation}

\subsection{Choice of Parameters}

\subsubsection{Number of Scales}
\label{number_of_scales}

In order to select the number of scales $J_0 \in \{1, \ldots, J\}$
of the wavelet transform decomposition,
the following criterion was introduced:~$J_0$
was chosen so that the coarsest approximation
scale
had a pseudo-frequency close to the
\egg dominant frequency $f_c$
of 4--6 cycles per minute for the canine subjects~\cite{Mintchev:2000}
and 3~cycles per minute for the humans.

The pseudo-frequency $f_{\text{pseudo}}$ of a given scale~$j$ is
\begin{equation}
f_{\text{pseudo}} = \frac{f_\psi}{j\cdot T_s}
,
\qquad j=1,2,\ldots,J,
\end{equation}
where~$T_s$ is the sampling period (0.1 \second) and~$f_\psi$
is the center frequency of a wavelet
(the frequency that maximizes the magnitude of the Fourier transform of the wavelet)~\cite{MisiMisi00}.
Consequently,
a scale~$J_0$ was selected which minimized the difference
$(f_{\text{pseudo}} - f_c)$.
Table~\ref{pseudo} shows the number of decomposition levels
for some common wavelets.

\renewcommand{\multirowsetup}{\centering}
\newlength{\LL}
\settowidth{\LL}{Wavelet}
\begin{table} %
\centering
\caption{Number of decomposition scales $J_0$ for some wavelets}
\label{pseudo}
\begin{tabular}{ccc}
\hline
\multirow{2}{\LL}{Wavelet} & \multicolumn{2}{c}{$J_0$} \\
\cline{2-3}
         & Canine & Human \\
\hline
Haar        & 7  &  8\\
Daubechies-2 & 6 & 7\\
Daubechies-3 & 7 & 7 \\
Coiflet-1 & 7 & 7 \\
\hline
\end{tabular}
\end{table}

\subsubsection{Compression Ratio}
\label{cr_section}

A compression ratio set $\CR \in \{3, 5, 7, 10 \}$
was selected and a matching procedure was carried out
aiming at optimizing the choice of wavelet.

\subsection{Optimization of the Wavelet Choice}

In the context of the present study,
a wavelet was sought that minimized
the $\prd$ between the original \egg signal and its reconstruction
for a given compression ratio.

However, the abundance of wavelets~\cite{Mallat:03}
makes such approach prohibitive.
As a result, some constraints on the choice of wavelet were introduced.

It is well known that wavelets can be generated from
discrete finite impulse response (FIR) filters~\cite{MisiMisi00}.
In the present work, the analysis was limited to
wavelets generated by FIR
filters with length no greater than six coefficients.
In this subset of wavelets one may find Haar, Daubechies-2, Daubechies-3, and Coiflet-1 wavelets,
to name the most popular ones~\cite{Mallat:03}.

This restriction is quite convenient, since all FIR filters of length up to six
that can be utilized to generate wavelets have simple parameterizations
of their coefficients~\cite{Zhou1993}.
For example, Pollen parameterization of 6-tap wavelet filters~\cite{Tew1992}
has two independent variables $(a,b)\in[-\pi,\pi]\times [-\pi,\pi]$.
Varying these two parameters,
a filter that generates a new wavelet can be defined.
Consequently, the Pollen parameterization defines a plane on which every
point is connected to a wavelet~\cite{Tew1992}.

Using the discussed compression scheme,
one can compute a~$\prd$ value for each wavelet
generated from a point with coordinates~$(a,b)$ on the
parameterization plane.
Doing so, a surface can be defined by the points~$(a,b,\prd)$.
Thus, the minima of this surface correspond to the
point coordinates~$(a,b)$ that generate a wavelet with good ``matching''
properties, since the $\prd$ values at these minima are small.

\begin{figure*}
\centering
\subfigure[$\CR =3$]{\epsfig{file=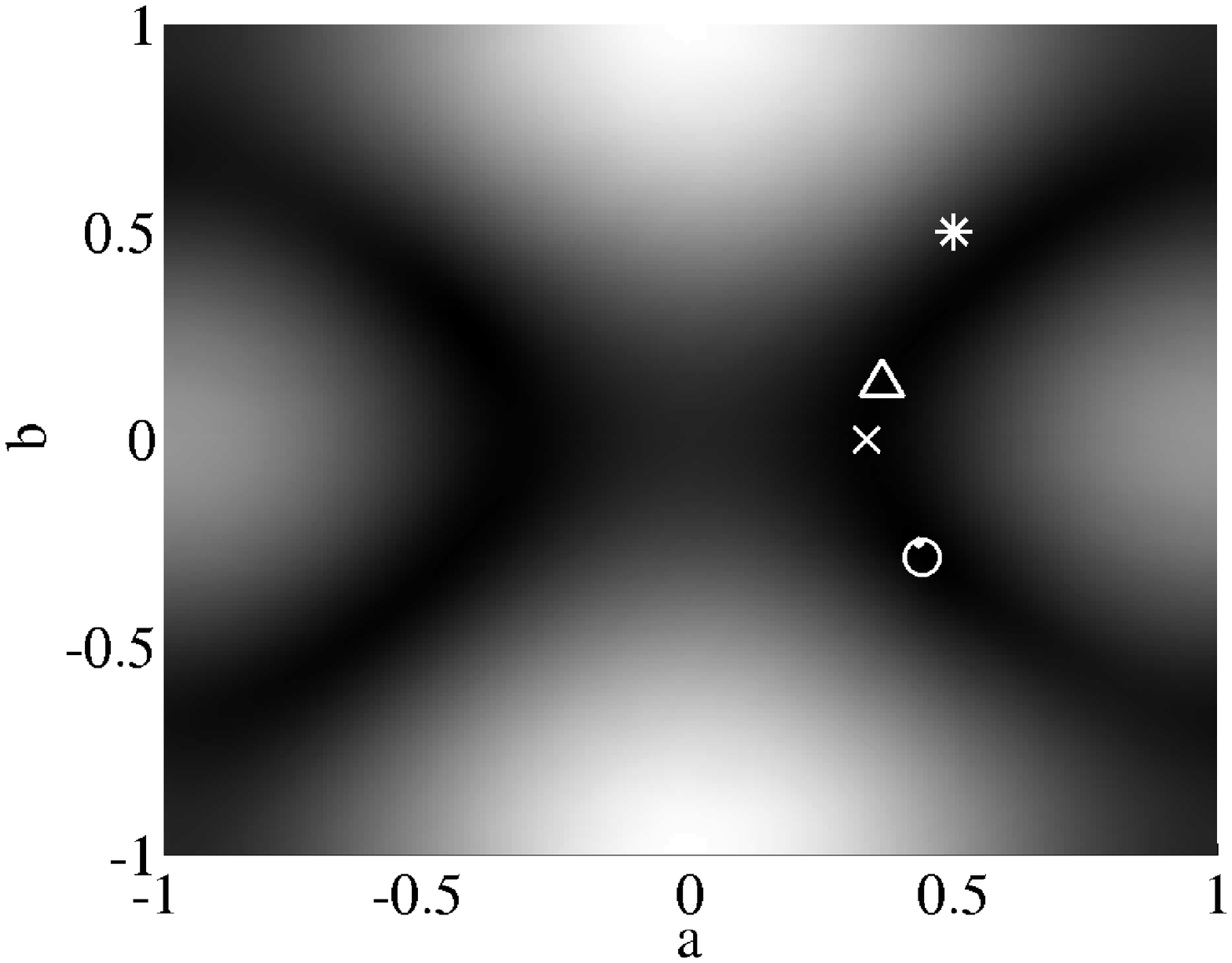,width=.45\linewidth}}
\subfigure[$\CR =10$]{\epsfig{file=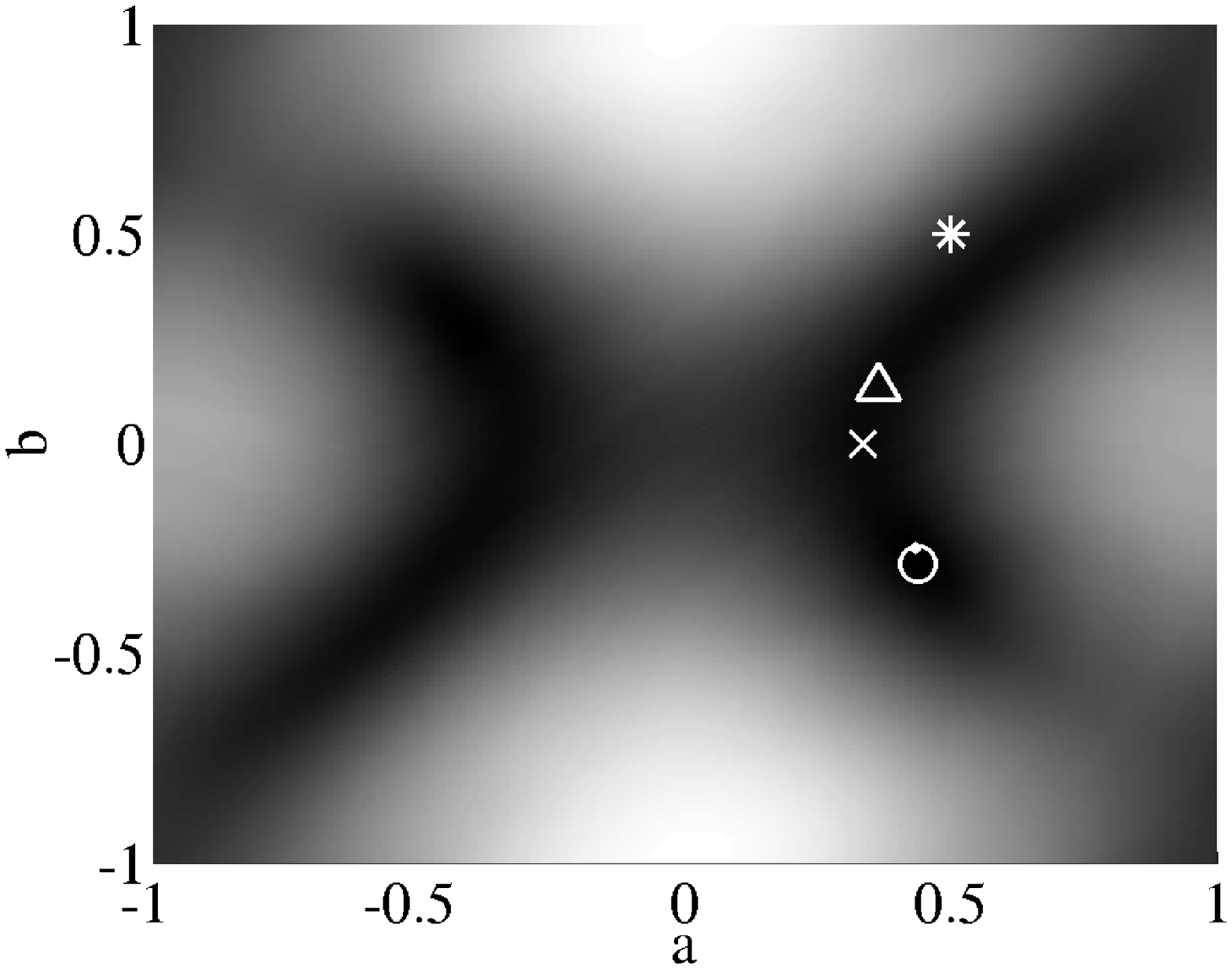,width=.45\linewidth}}
\subfigure[$\CR =3$]{\epsfig{file=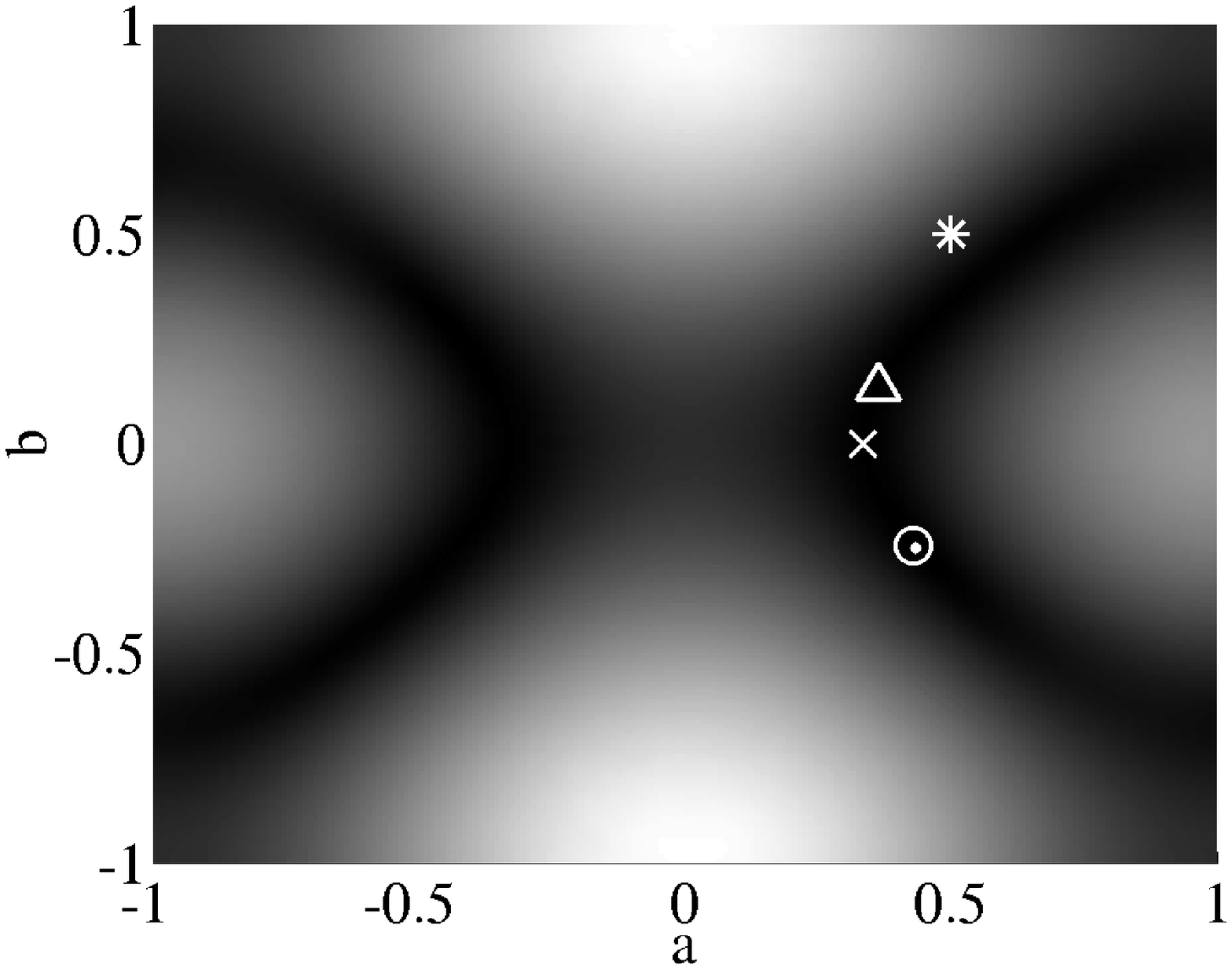,width=.45\linewidth}}
\subfigure[$\CR =10$]{\epsfig{file=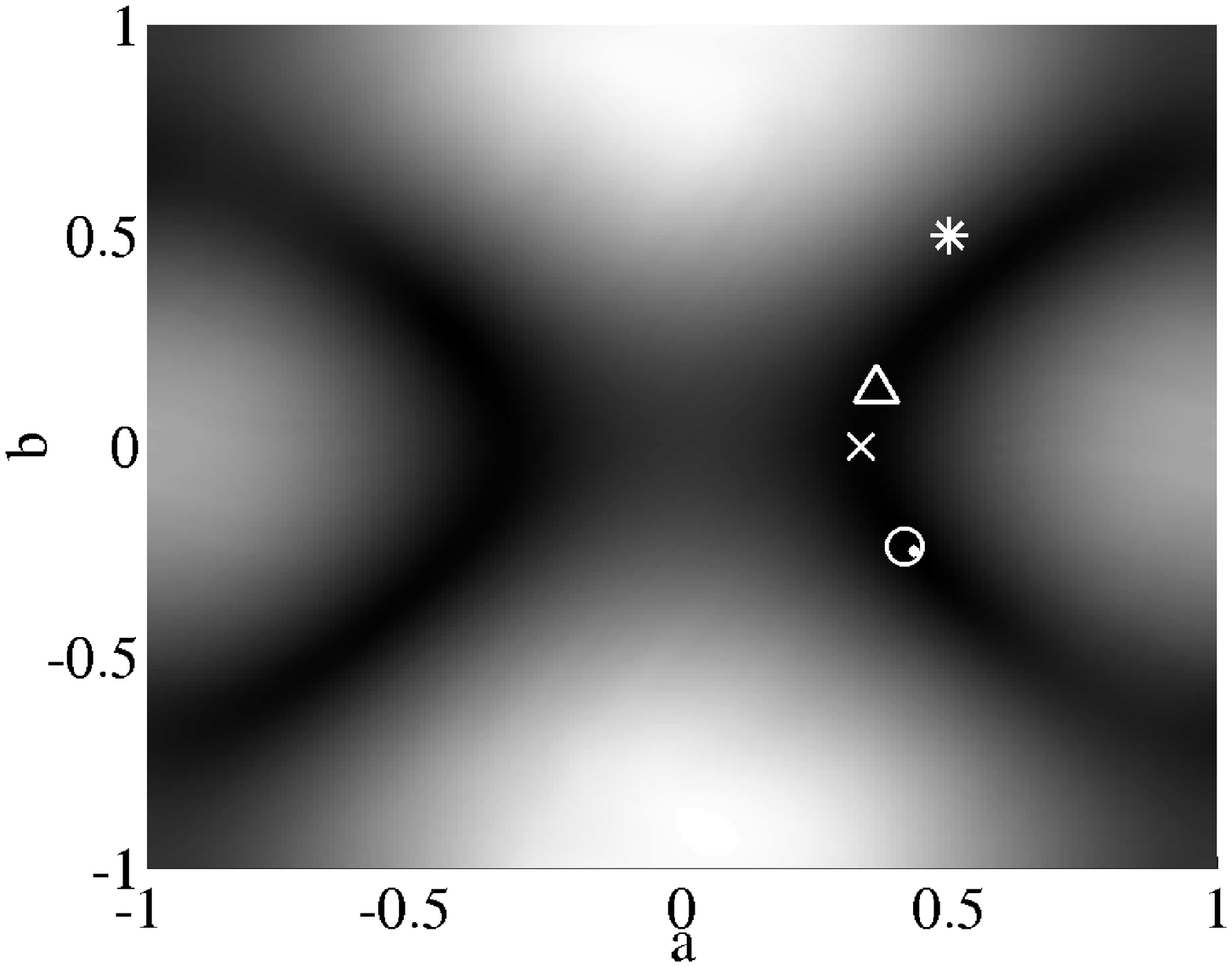,width=.45\linewidth}}
\caption{Plots generated after computing the $\prd$
surface for all possible wavelets on the parameterization plane,
using a canine \egg signal with $\CR=3$ (a), and $\CR=10$ (b).
A representative human \egg signal was used
to build $\prd$ surfaces with $\CR=3$ (c), and $\CR=10$ (d).
The minimum value is depicted by a circle ($\circ$).
The coordinate points
that correspond to Haar ($\ast$), Daubechies-2 ($\times$), Daubechies-3 ($\bullet$),
and
Coiflet-1 ($\triangle$) wavelets are shown.
The axes are normalized by~$\pi$.}
\label{prd-plane}
\end{figure*}

As a result, a set of point coordinates $(a_i, b_i)$
could be determined
on the parameterization plane, which minimizes the
$\prd$ for each \egg recording~$i$.
Fig.~\ref{prd-plane} shows typical surfaces
generated for basal canine and human \egg signals.

Taking the mean value of the minima,
the best wavelet parameterization $(a^\ast, b^\ast)$ could be defined.
Thus,
$(a^\ast,b^\ast)$
generates a wavelet that %
on average
``matches'' best the normal \egg recordings.

\section{Results}

\settowidth{\LL}{$\CR$}
\begin{table}
\caption{Optimal values of the parameterization}
\label{tab_results}
\begin{minipage}{\linewidth}
\begin{center}
\begin{tabular}{ccccc}
\toprule
\multirow{2}{\LL}{$\CR$} &
\multicolumn{2}{c}{Canine\footnote{Based on 16 canine subjects.}} &
\multicolumn{2}{c}{Human\footnote{Based on 6 volunteers.}} \\
\cmidrule{2-5}
 & $a^\ast$ &  $b^\ast$ &  $a^\ast$ &  $b^\ast$ \\
\midrule
3 & 0.4329 & $-$0.2608 & 0.3976 & $-$0.2335 \\
5 & 0.4323 & $-$0.2638 & 0.4293 & $-$0.2550 \\
7 & 0.4323 & $-$0.2700 & 0.4276 & $-$0.2551 \\
10 & 0.4293 & $-$0.2736 & 0.4279 & $-$0.2600 \\
\bottomrule
\end{tabular}
\end{center}
\end{minipage}
\end{table}

\begin{figure}
\centering
\epsfig{file=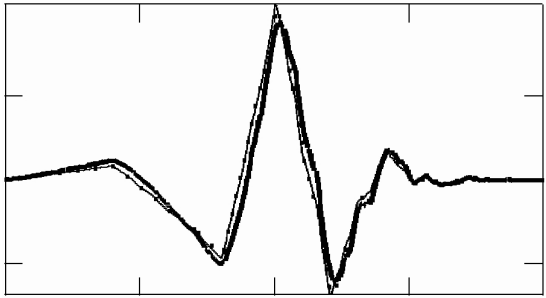,width=0.8\linewidth}
\caption{Comparison between the obtained optimal wavelets
for canine (dotted curve) and human (solid bold curve) \egg signals
and
the standard Daubechies-3 wavelet (solid thin curve)
using the
optimal parameterization %
for compression ratio of~3.
The similarity between Daubechies-3 and the optimal EGG wavelets is clearly evident.}
\label{best_wavelet}
\end{figure}

The optimal values of the wavelet
parameterization were determined
for the selected compression ratios
(Table~\ref{tab_results}).
The values for compression ratio of~3 were associated
with
the wavelets depicted in Fig.~\ref{best_wavelet}.
It is worth mentioning that the proposed wavelets were
very similar to the classic Daubechies-3 wavelet.
The correlation coefficient between Daubechies-3 and
the optimal EGG wavelet was 0.996 and 0.965, for canine and human
signals respectively.
Since Daubechies-3 is
(i)~very similar to the proposed wavelet;
and
(ii)~easily available in many software packages
(e.g., {\sc Matlab} (The Mathworks, Inc., Natick, MA, USA)),
it can be chosen instead of the proposed wavelets.
Previous empirical findings~\cite{Ryu:02} confirm this observation.

\section{Conclusion}

The problem of finding optimal wavelets to ``match''
\egg signals
was quantitatively addressed.
The proposed wavelets can be considered as tools to
further \egg signal analysis.
Moreover, the suggested methodology opens an avenue
towards the classification of
electrogastrograms based on the $\prd$ value of their wavelet compressed version,
either applying
the obtained optimal wavelets or the standard Daubechies-3.

\section*{Acknowledgments}

This work was partially supported by 
Natural Sciences and Engineering Research Council
of Canada (NSERC) and the National Council for Scientific and
Technological Development (CNPq, Brazil).

{\small
\bibliographystyle{IEEEtran}
\bibliography{ref}
}

\end{document}

%% file: electrode_table_gea.tex
{ \footnotesize
\begin{tabular}{cc}
\toprule
Channel &  \parbox{2cm}{\centering Electrode Combination} \\
\midrule
1      & a--b \\
2      & c--d \\
3      & e--f \\
4      & g--h \\
5      & i--j \\
6      & k--l \\
\bottomrule
\end{tabular}}

%% file: electrode_table_egg.tex
{ \footnotesize
\begin{tabular}{cc}
\toprule
Channel &  \parbox{2cm}{\centering Electrode Combination} \\
\midrule
7       & 1--2 \\
8       & 2--3 \\
9       & 3--4 \\
10      & 4--5 \\
11      & 1--3 \\
12      & 1--4 \\
13      & 2--5 \\
14      & 1--5 \\
\bottomrule
\end{tabular}}